\documentclass[amsmath,amssymb]{revtex4}
\usepackage{graphicx}
\usepackage{dcolumn}
\usepackage{bm}

\begin{document}

\title{Nonequilibrium Green function modelling of transport in
mesoscopic systems\footnote{Based on a talk presented at the
conference ``Progress in Nonequilibrium Green Functions, Dresden,
Germany, 19.-22. August 2002''}}

\author{Antti-Pekka Jauho}

\affiliation{Mikroelektronik Centret \\
Technical University of Denmark, Bldg. 345East, {\O}rsteds Plads, \\
DK-2800 Kgs. Lyngby, Denmark\\
E-mail: antti@mic.dtu.dk}


\begin{abstract}
A generalized Landauer formula, derived with the methods due to
Keldysh, and Baym and Kadanoff, is gaining widespread use in the
modeling of transport in a large number of different mesoscopic
systems.  We review some of the recent developments, including
transport in semiconductor superlattices, calculation of noise,
and nanoelectromechanical systems.
\end{abstract}

\maketitle

\section{Introduction}\label{sec:intro}
\subsection{Some background, and a few historical remarks}\label{subsec:background}

A mesoscopic\index{Mesoscopic device} transport measurement is
often concerned with a situation where the "device", whose
properties are the subject of the investigation, is connected to
structureless "contacts" via "ideal leads", i.e., in a situation
which is accessible via the Landauer formula (or, more generally,
the scattering approach to transport).  An important feature is
the fact that the size of the device is finite, and comparable to
other important length-scales of the system, such as the
phase-braking length\index{Phase-braking length} or impurity mean
free path.\index{Impurity mean free path} Thus the wave-properties
of the charge carriers are important, leading to a number of
interesting interference effects, such as weak
localization\index{Weak localization}, or universal conductance
fluctuations\index{Universal conductance fluctuations}. The
conductance $g$ of the device, for example, is then given by the
celebrated Landauer formula\cite{Land57,Land70}\index{Landauer
formula}
\begin{equation}
g=\frac{2e^2}{\hbar}|t|^2, \label{Landauer}
\end{equation}
where $t$ is the quantum mechanical transmission amplitude through
the device.

The expression given above holds for the one-channel case, for low
applied voltages, and in a situation when the "device" can be
modelled by a noninteracting system.  It is then natural to ask:
Can this equation be generalized to the case, when interactions
are important?  Or: Can this equation be extended to
time-dependent situations, or situations where superconductivity
or magnetism are essential for the physics?  How about strong
driving fields?

Many authors have addressed these questions, with a number of
different approaches.  In the spirit of this conference, this
review is concerned with the subset of these theories which use
the nonequilibrium Green function technique.  The earliest
applications to mesoscopic  transport, to my knowledge, are due to
French researchers\cite{Caro71a,Caro71b,Caro72,Comb71}:  these
researchers were mainly interested in inelastic effects in
tunneling through oxide barriers.  It is a curious side-note to
observe that these early, and pioneering, papers were essentially
forgotten during the 80's, but have obtained a substantial revival
since mid-90's, and are presently cited more often than ever
earlier. The explanation lies perhaps in the fact that the whole
idea of mesoscopics is newer than these early papers, and it took
the mesoscopic community a few years to realize the applicability
of these ideas.

For the purposes of the present review, the next important
development was the paper by Meir and Wingreen\cite{Meir92}, which
gave a very useful formal expression for the current in terms of
the {\it exact} Green function of the device (or "central
region"). This formula, and similar expressions obtained by other
groups, were then applied to the Kondo problem\index{Kondo
problem} out of equilibrium, a notoriously difficult problem,
which remains a topic of active research even today.  My previous
review\cite{Jauh00} in the first meeting of the present series
focused in some of these issues, paying particular attention to
the time-dependent generalization of the Meir-Wingreen
expression\cite{Jauh94}.  I shall not repeat any of that material,
but rather focus on other topics: the selection criterion has been
that either they were not discussed during the first meeting, or
that they are strictly post-1999 vintage. I have chosen to discuss
three examples: (i) Transport in a semiconductor
superlattice\index{Semiconductor superlattice}; (ii) Calculation
of the noise\index{Noise} in a spintronic
system\index{Spintronics}; and (iii) Tunneling transport in a
nanoelectromechanical (NEMS)\index{Nanoelectromechanics} device.
These three topics have a common feature: they all have practical
applications, and, dare I say in the present meeting, even
commercial potential.

It should be noted that there are many other recent applications
of the NGF to mesoscopic transport which, due to space and time
limitations, the present review does not address. One such example
is transport in nanowires, fabricated either with scanning
tunnelling microscope\index{Scanning tunneling microscope}, or
with break junctions\index{Break junctions}, which is a large
research field where many theoretical calculations emply the NGF
techniques.  An exhaustive review has recently become
available\cite{Agra02}. Molecular electronics\index{Molecular
electronics} holds enormous potential, and here a combination of
{\it ab initio} electronic structure calculations (within the
density-functional scheme)\index{Density-functional theory} and
NGF appears to the most promising theoretical
technique\cite{Paul02,Bran02}. Yet another example is the "circuit
theory" developed by Nazarov\cite{Naza94,Naza99}, which has
successfully been applied to a number of hybrid structures,
consisting of superconductors, ferromagnets, or semiconductors.
The more abstract field-theoretic formulations, based on
path-integrals and/or Grassmann variables fall also outside our
present purposes, even though they play an important role in the
study of disordered systems, or dephasing due to the environment.

\subsection{The basic equations, and their limitations}

For completeness, we sketch here a derivation of the basic
expressions used in the theory.  Several more complete accounts
are available elsewhere\cite{Jauh00,Jauh94,Haug96}.  A brief
reminder of how the nonequilibrium formalism works in the context
of mesoscopic transport measurements is also in place. One reasons
as follows. In the remote past the contacts and the central region
(i.e., the "device") are assumed to be decoupled, and each region
is in thermal equilibrium.  The equilibrium distribution functions
for the three regions are characterized by their respective
chemical potentials; these do not have to coincide nor are the
differences between the chemical potentials necessarily small. The
couplings between the different regions are then established and
treated as perturbations via  the standard techniques of
perturbation theory. The nonequilibrium nature of the problem
manifests itself in that symmetry of remote past and remote future
has been broken, and thus one must do the calculations on the
two-branch time contour. It is important to notice that the
couplings do not have to be small, e.g., with respect level to
spacings or $k_{\rm{B}} T$, and typically must be treated to all
orders.

Let us next consider some generic Hamiltonians:
$H=H_L+H_R+H_T+H_{\rm cen}$, or, explicitly:
\begin{eqnarray}
H &=& \sum_{k,\alpha\in L/R} \epsilon_{k,\alpha}
c^\dagger_{k,\alpha} c_{k,\alpha}\nonumber\\
&\quad& + \sum_{k,\alpha\in L/R;n} \left[V_{k\alpha;n}
c^\dagger_{k,\alpha} d_n + {\rm h.c.}\right] + H_{\rm
cen}\left[\{d_n\},\{d^\dagger_n\}\right]\;,
\end{eqnarray}
where the central part Hamiltonian must be chosen according to the
system under consideration.  The operators
$\{d_n\},\{d^\dagger_n\}$ refer to a complete set of
single-particle states of the central region. Occasionally we
specify explicitly the orbital and spin quantum numbers:
$n=m,\sigma$, and analogously for the states in the leads. The
derivation of the basic formula for the current does not require
an explicit form for $H_{\rm cen}$; the actual evaluation of the
formula of course requires this information.  We write $H_{\rm
cen}=\sum_n \epsilon_n d^\dagger_n d_n + H_{\rm int}$, where
$H_{\rm int}$ could be electron-phonon interaction,
\begin{equation}
H_{\rm int}^{\rm el-ph} = \sum_{m\sigma} d^{\dagger}_{m,\sigma}
d_{m,\sigma} \sum _{\bf q} M_{m,\bf q} \left[a^\dagger_{\bf q} +
a_{\bf q}\right]\;,\label{Helph}
\end{equation}
or an Anderson impurity:
\begin{equation}
H_{\rm int}^A = U \sum_m d_{m,\uparrow}^\dagger d_{m,\uparrow}
d_{m,\downarrow}^\dagger d_{m,\downarrow} \label{HAnd}\;.
\end{equation}
The current operator for the (say) left lead is
\begin{eqnarray}
I_L&=&-e\dot N_L=-\frac{ie}{\hbar}\left[H,N_L\right]\nonumber\\
&=&-\frac{ie}{\hbar}\sum_{k,n}\left[-V_{kL;n}c^\dagger_{kL}d_n+
V^*_{kL;n}d^\dagger_nc_{kL}\right]\;.\label{currop}
\end{eqnarray}
The physically relevant observables can be expressed in terms of
expectation values of the current operator, or its higher powers.
For example, one can show\cite{Jauh94,Haug96} that the current
leaving the left contact is
\begin{eqnarray}
\langle I_L \rangle=J_L(t) &=& - \frac{2e}{\hbar} \int_{-\infty}^t
dt_1 \int \frac{d\epsilon}{2\pi} {\rm ImTr} \Big \{
e^{-i\epsilon(t_1-t)} {\bf \Gamma}^L(\epsilon,t_1,t)
\nonumber\\
&\quad&\quad\times \left[ {\bf G}^<(t,t_1) + f^0_L(\epsilon) {\bf
G}^r(t,t_1)\right] \Big \}\;. \label{jtime}
\end{eqnarray}
Here the Green functions are defined by
\begin{eqnarray}
G_{nm}^<(t,t_1)&=&i\langle d_m^\dagger(t_1) d_n(t)\rangle\label{Gcorr}\\
G_{nm}^r(t,t_1)&=&-i\theta(t-t_1)\langle
[d_n(t),d_m^\dagger(t_1)]\rangle\label{Gret}\;,
\end{eqnarray}
$\Gamma_{mn}$ describes the coupling between the central region
and the contacts, and $f_L^0(\epsilon)$ is the equilibrium
distribution function of the left contact.  In the dc-limit,
(\ref{jtime}) reduces to the result of Meir and
Wingreen\cite{Meir92}:
\begin{eqnarray}
J &=& \frac{ie}{2\hbar} \int \frac{d\epsilon}{2\pi} {\rm Tr}\Big
\{ \left[ {\bf \Gamma}^L(\epsilon) - {\bf
\Gamma}^R(\epsilon)\right]
{\bf G}^<(\epsilon)\nonumber\\
&\quad&\quad + \left[ f_L^0(\epsilon){\bf \Gamma}^L(\epsilon) -
f_R^0(\epsilon) {\bf \Gamma}^R(\epsilon)\right]
\left[ {\bf G}^r(\epsilon) - {\bf G}^a(\epsilon)\right]\Big\}\\
&=&  \frac{i e}{\hbar}\int\frac{d\varepsilon}{2\pi}
\left[f_L(\varepsilon)-f_R(\varepsilon)\right] T(\varepsilon)\;,
\label{jprop}
\end{eqnarray}
where
\begin{equation}
T(\varepsilon)  = {\rm {Tr}} \left\lbrace \frac{{\bf
\Gamma}^L(\varepsilon){\bf \Gamma}^R(\varepsilon)}{{\bf
\Gamma}^L(\varepsilon)+{\bf\Gamma}^R(\varepsilon)} \bigl[{\bf
G}^r(\varepsilon)-{\bf  G}^a(\varepsilon)\bigr ]\right\rbrace \;.
\label{calT}
\end{equation}
The expressions (\ref{jtime}) and (\ref{jprop}) are the central
formal results whose consequences we explore in this
review.\footnote{In the Section 3 we also give an analogous
expression for the noise spectrum.}  They are formally exact, and
give the {\it tunneling} current for an interacting system coupled
to noninteracting contacts (or, more precisely, for contacts which
can be described by an effective single-body Hamiltonian).  Thus,
in a time-dependent situation the displacement
current\index{Displacement current} must be considered separately.
It should also be noted that these equations only {\it define} the
starting point of any calculation: to get into physical results
one must evaluate the correlation function and the
retarded/advanced Green function, Eqs.(\ref{Gcorr}) and
(\ref{Gret}), respectively.  These functions obey the Keldysh
equation,\index{Keldysh equation} and the (nonequilibrium) Dyson
equation:\index{Dyson equation}
\begin{eqnarray}
G^<&=&G^r\Sigma^<G^a,\label{Keldysh}\\
G^r&=&G^r_0+G^r_0\Sigma^rG^r\label{Dyson}.
\end{eqnarray}
The success of the theory depends on whether one can construct a
self-energy functional\index{Self-energy functional} that captures
the essential physics, and that a good solution can be found for
these coupled equations. Both of these steps may be hard indeed.

\section{Transport in a Semiconductor
Superlattice}\label{sec:superlattice}

In 1970 Esaki and Tsu\cite{Esak70} suggested that semiconductor
superlattices,\index{Semiconductor superlattice} man-made
structures which consist of alternating layers of different
semiconductor materials, would have physical properties which
could be used for a number of device applications.  Very shortly,
the spatial variations in the band-gaps will lead to a spatially
varying conduction band edge, which supports
minibands,\index{Miniband} which in turn diplay very interesting
transport properties, such as Bloch oscillations,\index{Bloch
oscillations} or negative differential resistance.\index{Negative
differential resistance}  A well known-result is the Esaki-Tsu
IV-characteristic\index{IV-curves},
\begin{equation}
I(V)=2I_{\rm max} V_0 \frac{V}{V^2+V_0^2},\label{Esaki}
\end{equation}
where $I_{\rm max}$ and $V_0$ depend on the physical paratmeres of
the system, such as the superlattice period, scattering rate, and
temperature. To derive expressions like this, three main
approaches have been used in the literature: (i) Miniband
transport\cite{Esak70},\index{Miniband transport} (ii)
Wannier-Stark hopping\cite{Tsu75},\index{Wannier-Stark hopping}
and (iii) sequential tunneling\cite{Mill94}.\index{Sequential
tunneling} The three different approaches have different domains
of validity, and are all likely to fail if the three basic energy
scales, i.e. scattering induced broadening, miniband width, and
potential drop per period all have comparable values. The basic
features of these three approaches are summarized in Figure 1.
 \vspace{0.3cm}

\begin{table}
\caption{The three standard approaches to miniband
transport, and the physical picture underlying them.(Courtesy of
A. Wacker.)}

\begin{tabular}{c|c|c|c|}
 & coupling $ T_1$ & field drop $eFd$ &
scattering $\Gamma$\\ \hline &&&\\
\begin{minipage}{4cm}
{Miniband conduction}\\
\includegraphics[width=4cm]{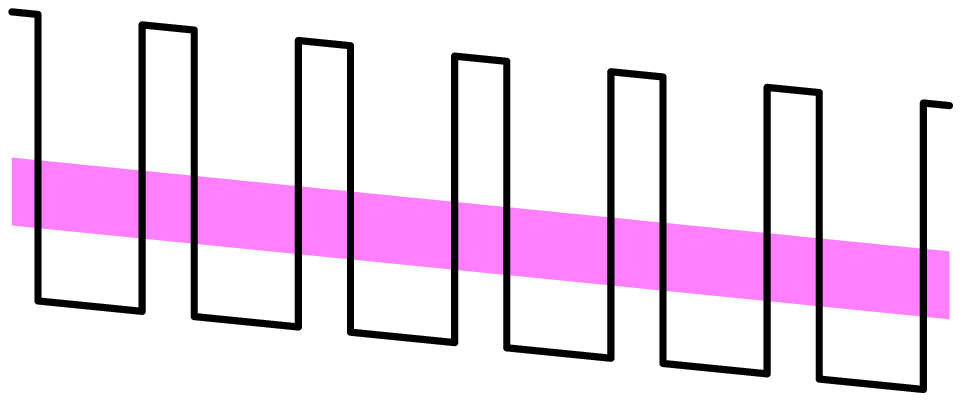}
\end{minipage}
&\begin{minipage}{2.0cm} exact\\
miniband\end{minipage} & acceleration &
golden rule \\ \hline & \multicolumn{2}{|c|}{ }&\\
\begin{minipage}{4cm}
{Wannier Stark\\ hopping}\\
\includegraphics[width=4cm]{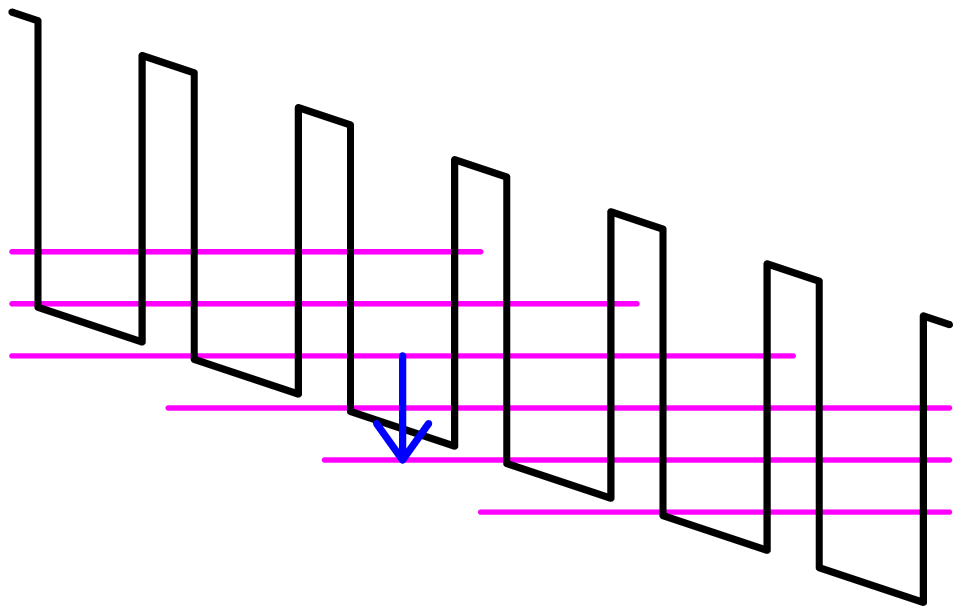}\end{minipage} &
 \multicolumn{2}{|c|}  {exact: Wannier Stark states} &
golden rule \\ \hline  &&&\\
\begin{minipage}{4cm}{Sequential tunneling}\\
\includegraphics[width=4cm]{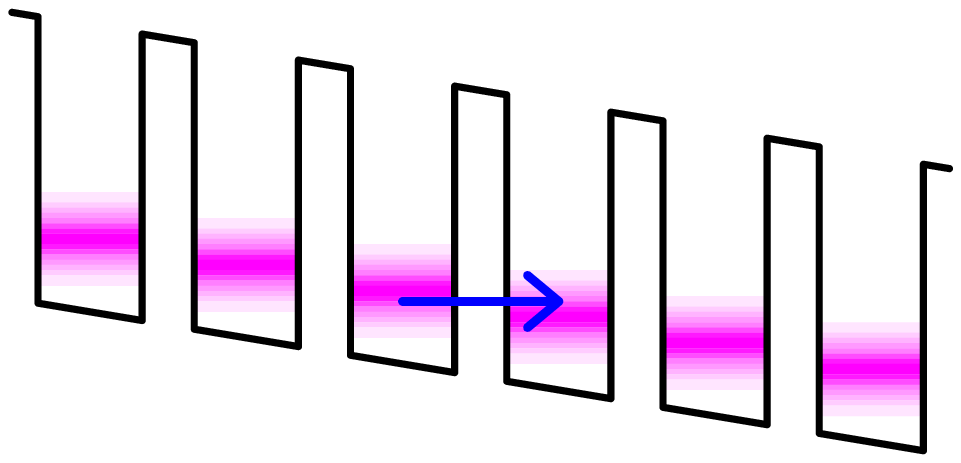}\end{minipage} &
 lowest order& \begin{minipage}{2.0cm} energy mismatch\end{minipage} &
\begin{minipage}{2.0cm} "exact"\\ spectral function
\end{minipage} \\ \hline
\end{tabular}
\end{table}
 \vspace{0.3cm}

In order to map out the boundaries of the various domains of
validity, and to access the region where the approaches (i)--(iii)
fail, a higher level theory is required. To achieve this, Andreas
Wacker and myself, together with several colleagues, launched a
program whose task was to develop a nonequilibrium Green function
theory for superlattice transport.\footnote{There are other
theoretical methods capable of including (some) quantum effects,
such as the density-matrix method\cite{Kuhn98,Bryk97}, or the
balance equation approach.\cite{Lei91}} Certain aspects of this
program are now completed\cite{Wack98,Wack99}, and in what follows
I will review some of the  highlights.  It should be noted that
the literature on superlattice transport is vast and here I can
give only a very superficial discussion; the reader is referred to
two recent review articles where a much fuller account can be
found\cite{Wack02,Boni02}.

Let me start with a few disclaimers. The quantum theory has {\it
not} yet been fully developed to the case when the electric field
is inhomogeneous (domain formation), nor is it available for the
time-dependent case (photo-assisted transport; progress is however
being made see, {\it e.g.}, Appendix C in the review by
Wacker\cite{Wack02}). For these important situations one has to
apply one of the simpler approaches discussed above. As far
scattering is concerned, impurity scattering and phonon scattering
have been discussed, but carrier-carrier interaction is still a
future task.

The task is now to solve the coupled Keldysh and Dyson equations,
Eq.(\ref{Keldysh}--\ref{Dyson}).  We adopt the tight-binding
representation\index{Tight-binding representation} of the
single-particle Hamiltonian:
\begin{equation}
H_{n,m}=(\delta_{nm-1}+\delta_{n,m+1})T_1+\delta_{n,m}(E_k-neFd),
\label{HTB}
\end{equation}
where $T_1$ is the nearest neighbor coupling, $E_k=\hbar^2
k^2/(2m)$ the kinetic energy perpendicular to the growth
direction, $F$ the applied field, and $d$ the superlattice period.
In this basis the Keldysh and Dyson equations\index{Keldysh
equation!tight-binding representation}\index{Dyson
equation!tight-binding representation}
(\ref{Keldysh}--\ref{Dyson}) read
\begin{eqnarray}
G^<_{mn}(E)&=&\sum_{m_1}G^r_{mm_1}\left(E+eFd\frac{m_1-n}{2}\right)\nonumber\\
&\times&\Sigma^<_{m_1}\left[E+eFd\left(m_1-\frac{m+n}
{2}\right)\right]\nonumber\\
&\times& G^a_{m_1n}\left(E+eFd\frac{m_1-m}{2}\right)\label{TBKeldysh}\\
G^r_{mn}(E)&=&g^r_m\left(E+eFd\frac{m-n}{2}\right)\nonumber\\
&\times&\left[\delta_{mn}+\sum_l
\Sigma^r_{ml}\left(E+eFd\frac{l-n}{2}\right)\right.\nonumber\\
&\quad&\times\quad
G^r_{ln}\left.\left(E+eFd\frac{l-m}{2}\right)\right]\;.\label{TBDyson}
\end{eqnarray}
Next one needs to specify the self-energies.  We have
considered\cite{Wack98,Wack99} impurity scattering, optical phonon
scattering, and mimicked acoustic phonon scattering\index{Phonon
scattering} by a very low-energy optical phonon, all in the
self-consistent Born approximation.\index{Self-consistent Born
approximation} By numerically solving these coupled equations,
computing the current, and comparing to the corresponding
IV-curves\index{IV-curves} found by the simpler approaches
(i)--(iii), we can construct a "phase-diagram" (see Figure 2),
which indicates where the simpler approaches hold, and where a
quantum approach is necessary.

\begin{figure}\label{fig:Regimes}
\includegraphics[width=8cm]{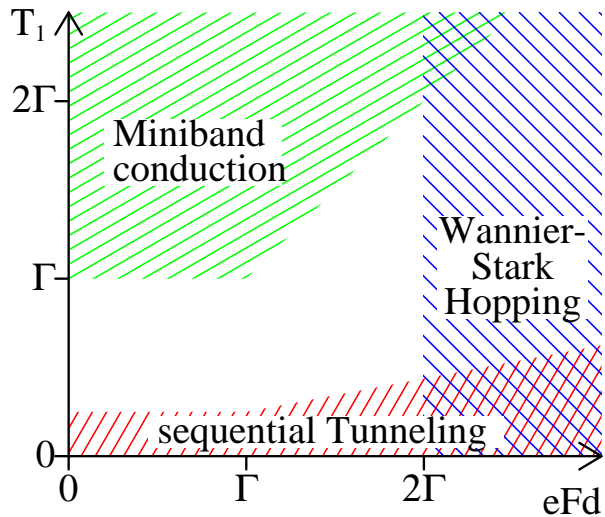} \caption{The range of validity
of various approaches to superlattice transport, in the parameter
space spanned by the nearest-neighbor coupling $T_1$, and the
potential energy drop $edF$ per period, in units of the scattering
width $\Gamma$.}
\end{figure}

We have also compared the quantum mechanical drift--velocity vs.
field relation\index{IV-curves} to the results obtained with a
semiclassical Monte Carlo simulation.\index{Monte Carlo
simulation} This is quite interesting because the two methods are
totally different, and both require computationally rather
intensive calculations. Typical results are shown in Figure 3. For
the parameters considered here, the Monte Carlo simulation gives
very good results, except that it misses the weak phonon replica
seen in the quantum calculation.

The approach sketched here is ideally suited to transport
phenomena where quantum phenomena, such as resonant
tunneling,\index{Resonant tunnelling} or phonon-assisted
tunneling\index{Phonon-assisted tunnelling}, play an important
role. Another application concerns quantum cascade
lasers\cite{Fais94}\index{Quantum cascade lasers}, where the
current injection occurs through a "funnel": the superlattice is
designed so that the miniband width varies with distance. Another
recent calculation concerns the evaluation of gain in such
structures\cite{Wack02b}.

\begin{figure}\label{fig:IV}
\includegraphics[width=10cm]{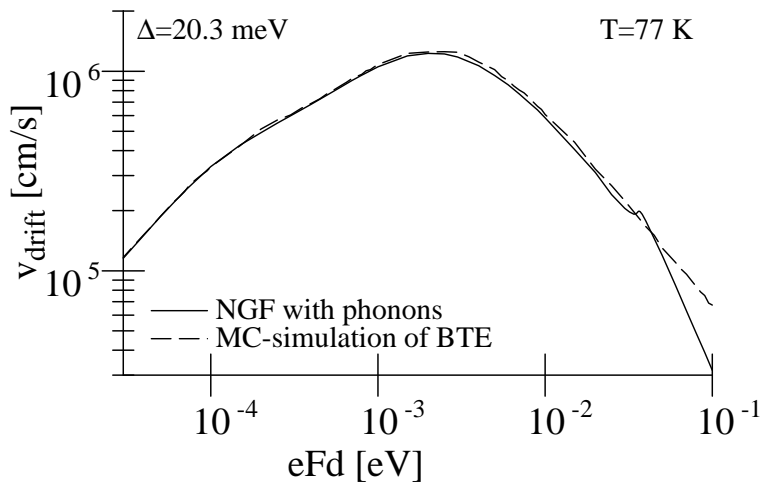} \caption{Drift velocity vs.
applied field.}
\end{figure}

\section{Noise in Spintronics}\label{sec:noise}

The emerging field of
spintronics\cite{dda02,saw01,gap98},\index{Spintronics} where, in
addition to the charge, also the electron spin is used to design
new devices, has led to fascinating and novel ideas such as spin
filters\cite{rf99,jce98,tk02}, spin field-effect
transistors\cite{sd90}, and  proposals for solid state quantum
computing\cite{dpv95}.\index{Quantum computing} For example,
quantum dot systems can in principle be used to control the
electron spin and are thus suitable for creating quantum bits
relevant for quantum gate\index{Quantum gates}
operations\cite{hae01}.

A detailed theoretical study of nonequilibrium transport
properties of spintronic devices is necessary in order to
understand the basic physical phenomena and to predict new
functionalities. Calculation of the current, for example, can give
the conductance/resistance of a system and its dependence on
magnetic field, Coulomb interaction, spin-flip and so on. On the
other hand, current fluctuations, due to the granularity of the
charge (shot noise\cite{ymb00})\index{Shot noise}, are also
relevant because their measurements can provide additional
information not contained in the average current\cite{clk97}.

Here we illustrate how the nonequilibrium Green function technique
can be used to calculate current and its fluctuations
(noise)\footnote{NGF calculations of noise have been performed by
several groups\cite{Chen91,Hers92,Ding97,Dong02}.} in a quantum
dot coupled to two ferromagnetic leads as a function of the
applied voltage for parallel (P) and antiparallel (AP)
lead-polarization alignments. We include Coulomb interaction in
the Hartree-Fock approximation as well as spin-flip in the dot. We
show that spin-flip makes the alignment of the lead polarizations
less important; both P and AP results coincide for large enough
spin-flip rates. This fact gives rise to a reduction of both Fano
factor\footnote{The Fano factor, defined as $\gamma=S(0)/2eI$,
characterizes deviations from the Poissonian
noise.\index{Poissonian noise}} \index{Fano factor} and tunnelling
magnetoresistance (TMR)\index{Tunneling magnetoresistance} as we
show below.

We model the central region with the  Hamiltonian
\begin{equation}
  H_D=\sum_{\sigma}\epsilon_0 d_\sigma^\dagger d_\sigma
  +Un_{\uparrow}n_{\downarrow}+R(d_{\uparrow}^\dagger
  d_{\downarrow}+d_{\downarrow}^\dagger d_{\uparrow}),
\label{HD}
\end{equation}
where $d_\sigma$ ($d_\sigma^\dagger$) destroys (creates) an
electron in the dot with spin $\sigma$ and the energy $\epsilon_0$
is spin independent\cite{wr01,pz02}. In addition, we assume that
the dot is a small enough in order to have only one active level
$\epsilon_0$. In the presence of a voltage the level shifts by
$\epsilon_0=\epsilon_{d}-\frac{eV}{2}$, where $\epsilon_{d}$ is
the dot level for zero bias.  In a more realistic calculation one
should determine the bias dependence self-consistently.  The
spin-flip scattering amplitude $R$ is viewed here as a
phenomenological parameter. The spin-flip process lifts the
degeneracy, splitting the quantum dot level to two states, let us
call them $\epsilon_{1,2}$, with corresponding operators.  The
current is readily evaluated with the formulas given in Section
1.2 with the result
\begin{equation}
J_L=\frac{2e}{\hbar}\mathrm{Re}\int dt_2 \mathrm{Tr}
  \{{\mathbf{G}}^r(t,t_2){\mathbf{\Sigma}}^{L <}(t_2,t)
  +{\mathbf{G}}^<(t,t_2){\mathbf{\Sigma}}^{L a}(t_2,t)\},
\label{Ieta}
\end{equation}
where $\mathbf{G}^r$ and $\mathbf{G}^<$ are the nonequilibrium dot
Green functions, with elements $G_{ij}^<(t,t_2)=i\langle
d_j^\dagger (t_2) d_i(t)\rangle$ and
$G_{ij}^r(t,t_2)=-i\theta(t-t_2)\langle\{d_i(t),d_j^\dagger(t_2)\}\rangle$.
The lesser (retarded, advanced) tunnelling self-energy is given by
\begin{eqnarray}
  \mathbf{\Sigma}^{L<(r,a)}&&(t_2,t)=\frac{1}{2}\sum_{k}|t_{kL}^2|\label{SelfL}\\
 &\times& \begin{pmatrix}
    g_{kL\uparrow}^{<(r,a)}(t_2,t)+g_{kL\downarrow}^{<(r,a)}(t_2,t) &
    g_{kL\uparrow}^{<(r,a)}(t_2,t)-g_{kL\downarrow}^{<(r,a)}(t_2,t) \\
    g_{kL\uparrow}^{<(r,a)}(t_2,t)-g_{kL\downarrow}^{<(r,a)}(t_2,t) &
    g_{kL\uparrow}^{<(r,a)}(t_2,t)+g_{kL\downarrow}^{<(r,a)}(t_2,t) \
  \end{pmatrix},
\nonumber
\end{eqnarray}
where $g_{kL\sigma}^{<(r,a)}$ is the lesser (retarded, advanced)
uncoupled Green function for lead $L$.  Equation (\ref{SelfL})
leads to a generalization of the coupling $\Gamma$ found in
Section 1.2 above; the coupling matrix now becomes
\begin{equation}\label{}
    \mathbf{\Gamma}^L=\frac{1}{2}
  \begin{pmatrix}
    \Gamma_{\uparrow}^L+\Gamma_\downarrow^L & \Gamma_\uparrow^L-
    \Gamma_\downarrow^L \\
    \Gamma_\uparrow^L-\Gamma_\downarrow^L & \Gamma_\uparrow^L
    +\Gamma_\downarrow^L \
  \end{pmatrix}.
\end{equation}
Accounting for Coulomb interaction in the Hartree-Fock
approximation, we can write down a matrix Dyson
equation\index{Dyson equation} for the retarded Green function,
$\mathbf{G}^r=\mathbf{G}^{0r}+\mathbf{G}^{0r}\mathbf{\Sigma}^r\mathbf{G}^r$,
and a Keldysh equation\index{Keldysh equation} for the lesser
Green function
$\mathbf{G}^<=\mathbf{G}^r\mathbf{\Sigma}^<\mathbf{G}^a$, where
$\mathbf{G}^{0r}$ is the uncoupled dot Green function. In these
equations the self energies are the sum of the left and right self
energies, i.e., $\Sigma^{(r,<)}=\Sigma^{L(r,<)}+\Sigma^{R(r,<)}$.
A self-consistent calculation is required to calculate $\langle
n_{\overline{i}}\rangle$ and $\langle d_{\overline{i}}^\dagger d_i
\rangle$, which are given by the lesser Green function, $\langle
d_{j}^\dagger d_i\rangle=\int \frac{d\omega}{2\pi}\mathrm{Im}
G_{ij}^<(\omega)$.

The current operator can be written as its average value plus some
fluctuation, i.e., $I_\eta=J_\eta+\delta{I}_\eta$ (here $\eta=L/R$
labels the contacts). In our system there are two sources of
noise, namely, thermal noise\index{Thermal noise} and shot
noise.\index{Shot noise} The first one is due to thermal
fluctuations in the occupations of the leads. It vanishes for zero
temperature, but can be finite for $T\neq 0$ and $eV=0$. On the
other hand, shot noise is due to the granularity of the electron
charge; it is a nonequilibrium property of the system in the sense
that it is nonzero only when there is a finite current ($eV\neq
0$). To calculate the noise (thermal+shot noise) we use the
definition $S_{\eta\eta'}(t-t')=\langle\{\delta{I}_\eta
(t),\delta{I}_{\eta'}(t')\}\rangle$, which can also be written as
$S_{\eta\eta'}(t-t')=\langle
\{{I}_\eta(t),{I}_{\eta'}(t')\}\rangle-2J_\eta^2$. After a lengthy
but straightforward calculation\cite{fms02}, we find for the noise
power spectrum ({\it dc} limit; a scalar version of this equation
has been found earlier\cite{Dong02})

\begin{equation}\label{}
\begin{split}
  S_{\eta\eta'}(0)=\frac{e^2}{\hbar}\int \frac{d\omega}{2\pi} &\{
    \delta_{\eta\eta'}in_\eta \mathbf{\Gamma}^\eta \mathbf{G}^>
    -\delta_{\eta\eta'}i(1-n_\eta)\mathbf{\Gamma}^\eta \mathbf{G}^<
    +\mathbf{G}^<\mathbf{\Gamma}^\eta \mathbf{G}^>\mathbf{\Gamma}^{\eta'}\\&
    -n_\eta(1-n_{\eta'})\mathbf{G}^r\mathbf{\Gamma}^\eta \mathbf{G}^r
    \mathbf{\Gamma}^{\eta'}-n_{\eta'}(1-n_\eta)\mathbf{G}^a \mathbf{\Gamma}^\eta
    \mathbf{G}^a \mathbf{\Gamma}^{\eta'}\\&
   - \mathbf{G}^<\mathbf{\Gamma}^\eta [(1-n_{\eta'})\mathbf{G}^r-
   (1-n_\eta)\mathbf{G}^a]\mathbf{\Gamma}^{\eta'}\\&
   +(n_\eta
    \mathbf{G}^r-n_{\eta'}\mathbf{G}^a)\mathbf{\Gamma}^\eta \mathbf{G}^>
    \mathbf{\Gamma}^{\eta'}\}.\label{Setaeta}
\end{split}
\end{equation}
The $dc$ noise (zero frequency) is position independent, and it is
possible to show that
$S_{LL}(0)=S_{RR}(0)=-S_{LR}(0)=-S_{RL}(0)$\cite{ymb00}.  In our
numerics we make a few simplifying assumptions.  We assume that
the couplings $\mathbf{\Gamma}^\eta$ are energy independent, but
allow a polarization dependence.  For the physical parameters we
use accepted values from the current literature\cite{wr01}.  Our
Hartree-Fock approximation for the electron-electron interaction
does not include correlations of the Kondo type, however we do not
expect these to change our results in the present range of
parameters.

\begin{figure}[h]
\begin{center}
\includegraphics[width=7cm]{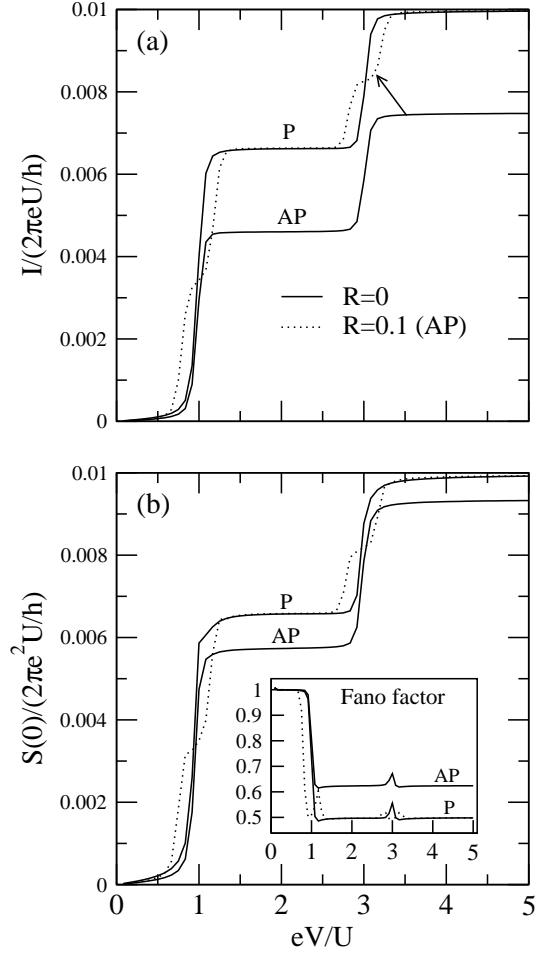}
\end{center}
\caption{Current and noise as a function of the bias for parallel
(P) and antiparallel (AP) alignments and with $R=0$ and $0.1U$.
The curves for $R=0.1U$ are only for the AP alignment; observe
that these are almost on top of the P curves, except within the
sloping region around $U$ and $3U$. Both current and noise are
reduced when the right lead changes its polarization from P to AP,
following the typical behavior of TMR. The inset shows a
suppression of the AP-Fano factor due to spin-flip.} \label{fig4}
\end{figure}

Figure 4 shows current (a) and noise (b) as a function of the bias
with $R=0$ (solid line) and $R=0.1U$ (dotted line) for both P and
AP configurations. Because P and AP curves for $R=0.1U$ coincide,
we plotted only the AP case. The first enhancement of the current
and noise at $eV=U$ happens when $\epsilon_0$ crosses the left
chemical potential, allowing electrons to tunnel from the emitter
(left lead) to the dot and then to the collector (right lead). The
current and noise remain constant until the second level
$\epsilon_0+U$ reaches $\mu_L$ at $eV=3U$, when another
enhancement is observed. Each enhancement corresponds to a peak in
the differential conductance $\sigma_{\rm diff}$. When the system
changes from parallel (P) to antiparallel (AP) configurations the
current is reduced. This is a typical behavior of tunnelling
magnetoresistance (TMR)\index{Tunneling magnetoresistance}. The
noise is also affected by this resistance variation, showing a
similar reduction.

Looking at the effects of spin-flip on current and noise we see
that the AP curves with $R=0.1U$ (dotted lines) tend to lie on the
P curves with $R=0$, thus showing that lead alignments are less
important when spin-flip plays a part. This AP current enhancement
due to spin-flip gives rise to a reduction of the TMR; since ${\rm
TMR}=(I_P-I_{AP})/I_{AP}$, when $I_{AP}\rightarrow I_P$ we have
${\rm TMR}\rightarrow 0$. For a somewhat simpler model W.
Rudzi{\'n}ski {\it et al.}\cite{wr01} found a similar behavior.

\section{Nanoelectromechanical Systems}\label{sec:nems}

Microelectromechanical systems (MEMS)\index{Microelectromechanical
systems} are today an important part of our technology.  Their
functionality is based on combining mechanical and electronic
degrees of freedom, the great advantage being that the whole
device can be fabricated with standard Si-processing technology.
Typical applications include hearing aids, sensors, or actuators.
As the fabrication technology gets refined, we soon expect to find
systems where the mechanical parts are in the nanometer range, see
the review by Craighead\cite{Crai00}; hence the acronym
NEMS.\index{NEMS} An example is the nanomechanical electron
shuttle constructed by Erbe {\it et al.}\cite{Erbe01} (see Figure
5), based on the theoretical ideas of Gorelik {\it et
al.}.\cite{Gore98}
\begin{figure}[h]
\begin{center}
\includegraphics[width=7cm]{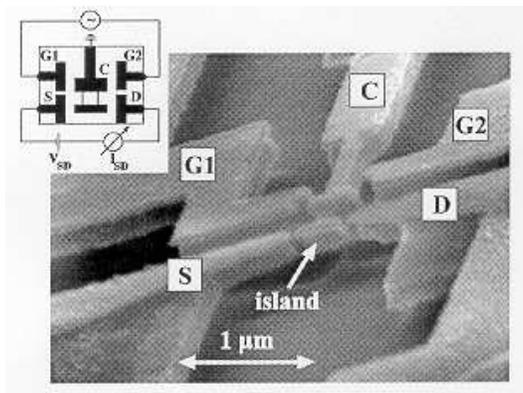}
\end{center}
\caption{The nanomechanical electron shuttle.  An ac-voltage
coupled to the gates $G_1$ and $G_2$ causes the central electrode,
the "clapper", $C$ to oscillate, and charges tunnel from the
source $S$ to the island, and from the island to the drain $D$.}
\label{fig5}
\end{figure}
Transport through the device was modelled with
rate-equations\cite{Erbe01}, however these are not expected to
work when one enters the coherent transport regime, at lower
temperatures and smaller devices, and a quantum theory of
transport must be invoked.  Such an approach has recently been
formulated by Fedorets {\it et al.}\cite{Fedo02}, and I'll give a
brief introduction to this topic.

The archetypal NEMS device consists of a moving part, and
electrodes.  We shall model the moving part by a quantum well,
whose coupling to the electrodes are position dependent: the
tunnelling amplitude is written as
$T_{L/R}(x(t))=\tau_{L/R}\exp[\mp x(t)/\lambda]$, where $\lambda$
is some characteristic tunnelling length, and $x(t)$ is the
center-of-mass coordinate of the moving quantum dot.  The
electronic degrees of freedom are governed by exactly the same
Hamiltonians as discussed in Section 2; now the dot-level depends
on the dot's location via $\epsilon_d=\epsilon_0-E x(t)$, where
$E$ is the electric field, which depends on the applied
bias.\footnote{In principle a rather complicated electrostatic
calculation, including the screening due to all neighboring
metallic bodies, is required.}  The center-of-mass of the dot
obeys Newton's equation of motion,
\begin{equation}
{\ddot x}+\omega_0^2 x = F(x)/M\;,\label{Newton}
\end{equation}
where $M$ is the mass of the grain, $\omega^2_0=k/M$ is the
characteristic frequency, and $F$ is the force, which includes
both the electric force acting on the charge(s) on the dot, and an
exchange force, which arises due to the $x$-dependence of the
tunnelling matrix elements.  One can evaluate the force via
$F=-\langle \partial H/\partial x \rangle$, and the result is
\begin{equation}
F(t)=-i E G^<(t,t)+2/\lambda \sum_{\alpha,k}(-1)^\alpha
T_\alpha(t){\rm Im}[G^<_{\alpha,k}(t,t)],\label{force}
\end{equation}
using the notation of Section 2.  The problem is thus (again)
reduced to the determination of the two lesser Green function:
$G^<$ for the dot, and the non-diagonal $G^<_{\alpha,k}=i\langle
c_{k,\alpha}^\dagger d(t)\rangle$. Fedorets {\it et
al.}\cite{Fedo02} calculate these for a noninteracting system, and
proceed to present an analysis of the mechanical stability of the
system: does the dot execute regular oscillations (whose
frequencies are determined by an appropriate linearization of
(\ref{Newton}) and (\ref{force}), or does it perhaps become
unstable, as the bias is increased?  The details of the analysis
are not our concern here; the upshot is that above a certain
threshold value an instability results.  The analysis has also
bearing on a recent experiment\cite{Park00}, where vibronic
anomalies were observed in a single-${\rm C}_{60}$-transistor when
current was passed through it.

The analysis of Fedorets {\it et al.}\cite{Fedo02} is very
interesting and suggests for several further refinements.  For
example, what are the results for an interacting system (Coulomb
blockade)? How about the environmental degrees of freedom? This
issue was addressed recently Armour and MacKinnon\cite{Armo02}, in
a slightly different context.  Finally, is it possible to combine
the spintronic effects with charge shuttles?  Can one envisage a
spin shuttle?  Will there be a new technology called NEMSS
(nanoelectromechanical spin systems)?\index{NEMSS}

\section{Conclusions}
I have reviewed some of the post-1999 developments in applying NGF
to modelling of transport in mesoscopic systems.  The common theme
in my review has been to focus on "real devices", which may have
"real applications".  I find it very pleasing that the NGF
technique, often regarded as an academic exercise most suited for
theoretical games, is now becoming a strong tool in the analysis
of practical devices. This trend is also confirmed by several
other talks at this conference.  At the same time there is still
much room for theoretical refinements, and I'm convinced that in
the coming years we will witness significant progress in this
field, both abstract and practical.

\section*{Acknowledgments}
The author is grateful to acknowledge fruitful collaborations with
A. Wacker, and F. Macedo and J. C. Egues, which resulted in the
material presented in Sections 2 and 3, respectively.




\begin{thebibliography}{0}

\bibitem{Land57}
R. Landauer,  IBM J. Res. Dev. {\bf 1}, 233 (1957).

\bibitem{Land70}
R. Landauer, Philos. Mag. {\bf 21}, 863 (1970).


\bibitem{Caro71a}
C. Caroli, R. Combescot, D. Lederer, P. Nozieres , and D.
Saint-James, J. Phys. C {\bf 4}, 2598 (1971).

\bibitem{Caro71b}
C. Caroli, R. Combescot, P. Nozieres, and D. Saint-James, J. Phys.
C {\bf 4}, 916 (1971).

\bibitem{Caro72}
C. Caroli, R. Combescot, P. Nozieres, and D. Saint-James, J. Phys.
C. {\bf 5}, 21 (1972).

\bibitem{Comb71}
R. Combescot, J. Phys. C {\bf 4}, 2611 (1971).

\bibitem{Meir92}
Y. Meir and N. S. Wingreen,
Phys. Rev. Lett. {\bf 68}, 2512 (1992).

\bibitem{Jauh00}
A. P. Jauho, pp. 250 -- 273, in "Progress in Nonequilibrium
Green's Functions", Ed. M. Bonitz, World Scientific, Singapore
(2000).

\bibitem{Jauh94}
A. P. Jauho, N. S. Wingreen, and Y. Meir, Phys. Rev. B {\bf 50},
5528 (1994).

\bibitem{Agra02}
N. Agra\"{\i}t, A. Levy Yeati, and J. M. van Ruitenbeek, to appear
in Physics Reports (2002), archived at cond-mat/0208239.

\bibitem{Paul02}
M. Paulsson, F. Zahid, and S. Datta, to appear in "Nanoscience,
Engineering and Technology Handbook", Eds. W. Goddard, D. Brenner,
S. Lyshevski, and G. Iafrate, CRC Press (2002), archived at
cond-mat/0208183.

\bibitem{Bran02}
M. Brandbyge, J. L. Mozos, P. Ordej{\'o}n, J. Taylor, and K.
Stokbro, Phys. Rev. B {\bf 65}, 165401 (2002).

\bibitem{Naza94}
Yu. V. Nazarov, Phys. Rev. Lett. {\bf 73}, 1420 (1994).

\bibitem{Naza99}
Yu. V. Nazarov, Superlattices Microstruct. {\bf 25}, 1221 (1999).

\bibitem{Haug96}
H. Haug and A. P. Jauho, {\it Quantum Kinetics in Transport
and Optics of Semiconductors} (Springer Series in Solid-State
Sciences, Vol. 123, Springer-Verlag, Berlin Heidelberg, 1996).

\bibitem{Esak70}
L. Esaki and R. Tsu, IBM J. Res. Develop. {\bf 14}, 61 (1970).

\bibitem{Tsu75}
R. Tsu and G. D{\"o}hler, Phys. Rev. B {\bf 12}, 680 (1975).

\bibitem{Mill94}
D. Miller and B. Laikhtman, Phys. Rev. B {\bf 50}, 18426 (1994).

\bibitem{Kuhn98}
T. Kuhn, pp. 173 -- 214, in: E. Sch{\"o}ll (Ed.), "Theory of
Transport Properties of Semiconductor Nanostructures", Chapman
$\&$ Hall, London, 1998.

\bibitem{Bryk97}
V. V. Bryksin and P. Kleinert, J. Phys.: Condens. Matt. {\bf 9},
15827 (1997).

\bibitem{Lei91}
X. L. Lei, N. J. M. Horing, and H. L. Cui, Phys. Rev. Lett. {\bf
66}, 3277 (1991).

\bibitem{Wack98}
A. Wacker and A. P. Jauho, Phys. Rev. Lett. {\bf 80}, 369 (1998).

\bibitem{Wack99}
A. Wacker, A. P. Jauho, S. Rott, A. Markus, P. Binder, and G. H.
D{\"o}hler, Phys. Rev. Lett. {\bf 83}, 836 (1999).

\bibitem{Wack02}
A. Wacker, Phys. Rep. {\bf 357}, 1 (2002).

\bibitem{Boni02}
L. L. Bonilla, J. Phys.: Condens. Matt. {\bf 14}, R341 (2002).

\bibitem{Fais94}
J. Faist, F. Capasso, D. L. Sivco, C. Sirtori, A. L. Hutchinson,
and A. Y. Cho, Science {\bf 24}, 553 (1994).

\bibitem{Wack02b}
A. Wacker, to appear in Phys. Rev. B (2002).

\bibitem{dda02}
D. D. Awschalon, M. E. Flatt{\'e} and N. Samarth, Sci. Am. {\bf
286}(6), 66 (2002).

\bibitem{saw01}
S. A. Wolf., D. D. Awschalom, R. A. Buhrman, J. M. Daughton, S.
von Moln{\'a}r, M. L. Roukes, A. Y. Chtchelkanova and D. M.
Treger, Science {\bf 294}, 1488 (2001).

\bibitem{gap98}
G. A. Prinz, Science {\bf 282}, 1660 (1998).

\bibitem{rf99}
R. Fiederling, M. Keim, G. Reuscher, W. Ossau, G. Schmidt, A. Waag
and L. W. Molenkamp, Nature {\bf 402}, 787 (1999).

\bibitem{jce98}
J. C. Egues, Phys. Rev. Lett. {\bf 80}, 4578 (1998).

\bibitem{tk02}
T. Koga, J. Nitta, H. Takayanagi and S. Datta, Phys. Rev. Lett.
{\bf 88}, 126601 (2002).

\bibitem{sd90}
S. Datta and B. Das, Appl. Phys. Lett. {\bf 56}, 665 (1990)

\bibitem{dpv95}
D. P. DiVincenzo, Science {\bf 270}, 255 (1995).

\bibitem{hae01}
H. A. Engel, P. Recher and D. Loss, Solid State Commun. {\bf 119},
229 (2001).

\bibitem{ymb00}
For a review on shot noise, see Ya. M. Blanter and M.
B{\"u}ttiker, Phys. Rep. {\bf 336}, 2 (2000).

\bibitem{clk97}
For an example in the context of fractional charge, see
C. L Kane and M. P. A. Fisher, Nature {\bf 389}, 119 (1997).

\bibitem{Chen91}
L. Y. Chen and C. S. Ting, Phys. Rev. B {\bf 43}, 4532 (1991).

\bibitem{Hers92}
S. Hershfield, Phys. Rev. B {\bf 46}, 7061 (1992).

\bibitem{Ding97}
G. H. Ding and T. K. Ng, Phys. Rev B {\bf 56} 15521 (1997).

\bibitem{Dong02}
B. Dong and X. L. Lei, J. Phys.: Condens. Matt. {\bf 14}, 4963
(2002).

\bibitem{wr01}
W. Rudzi{\'n}ski and J. Barna{\'s}, Phys. Rev. B {\bf 64}, 085318
(2001)

\bibitem{pz02}
P. Zhang, Q. K. Xue and X. C. Xie, cond-mat/0201465 (2002).

\bibitem{fms02}
F. M. Souza, J. C. Egues and A. P. Jauho, in preparation.

\bibitem{rs02}
R. {\'S}wirkowicz, J. Barna{\'s} and M. Wilczy{\'n}ski., J. Phys.:
Condens. Matt. {\bf 14}, 2011 (2002).

\bibitem{Crai00}
H. G. Craighead, Science {\bf 290}, 1532 (2000).

\bibitem{Erbe01}
A. Erbe, C. Weiss, W. Zwerger, and R. H. Blick, Phys. Rev. Lett.
{\bf 87}, 096106 (2001).

\bibitem{Gore98}
L. Y. Gorelik, A. Isacsson, M. B. Voinova, B. Kasemo, R. I.
Shekther, and M. Jonson, Phys. Rev. Lett. {\bf 80}, 4526 (1998).

\bibitem{Fedo02}
D. Fedorets, L. Y. Gorlik, R. I. Shekhter, and M. Jonson,
Europhys. Lett. {\bf 58}, 99 (2002).

\bibitem{Park00}
H. Park, J. Park, A. K. L. Lim, E. H. Anderson, Alivistos A. P.,
and P. L. McEuen, Nature {\bf 407}, 58 (2000).

\bibitem{Armo02}
A. D. Armour and A. MacKinnon Phys. Rev. B {\bf 66}, 035333
(2002).


\end{thebibliography}
\end{document}